\documentclass[final]{ias2}

\usepackage{graphicx} 
\usepackage{multirow}
\usepackage{array} 

\usepackage{hyperref} 
\begin{document}
	
	%\markboth{Pramana class file for \LaTeX 2e}{Xerxes Yu, et. al.}
	
	\title{First results of evaporation residue cross-section measurements of $^{32}$S+$^{208}$Pb system}

	\author[1,2]{R. Sariyal} 
	\email{ranjansariyal17@gmail.com}
	\author[2]{I. Mazumdar} 
	\email{indra@tifr.ac.in}
	\author[1]{D. Mehta}
	\author[4]{ N. Madhavan}
	\author[4]{ S. Nath}
	\author[4]{J. Gehlot}
	\author[4]{Gonika}
	\author[2]{S. M. Patel}
	\author[2]{P. B. Chavan}
	
	\author[5]{S. Panwar}
	\author[5]{V. Ranga}
	
	\author[6]{A. Parihari}
	
	\address[1]{Department of Physics, Panjab University, Chandigarh-160014, India.}
	\address[2]{Department of Nuclear and Atomic Physics, Tata Institute of Fundamental Research, Colaba-400005, Mumbai, India.}
	%\address[3]{Department of Applied Sciences, Chandigarh Engineering College-CGC, Mohali, Punjab-140307, India.}
	\address[4]{Inter-University Accelerator Centre, Aruna Asaf Ali Marg, New Delhi-110067, India.}
	\address[5]{Department of Physics, Indian Institute of Technology, Roorkee, Roorkee-247667, Uttarakhand, India.}
	\address[6]{Department of Physics and Astrophysics, University of Delhi, New Delhi-110007, India.}

	\begin{abstract}
		The dynamics of heavy ion-induced reactions play a critical role in forming super heavy elements (SHE), and one clear signature of the SHE formation is the evaporation residue (ER). In our pursuit of SHE, we present the heaviest element populated in India for ER cross-section measurements. These are the first-ever measurements of the Evaporation Residue (ER) cross-sections for the nuclear reactions between $^{32}$S and $^{208}$Pb. These measurements were conducted above the Coulomb barrier at four distinct beam energies in the laboratory frame, ranging from 176 to 191 MeV at the pelletron Linac facility at Inter-University Accelerator Centre (IUAC), New Delhi. The Hybrid Recoil Mass Analyzer (HYRA) in a gas-filled mode was employed  for these experiments. The obtained range of ER cross-sections enriches our knowledge and helps advance the field of heavy ion-induced reactions, especially in the context of super heavy element formation.
	\end{abstract}
	
	\keywords{Evaporation Residue (ER), Cross-section, Compound nucleus, Super Heavy Elements (SHE)}
	
	\pacs{Appropriate pacs here}

	\maketitle

%	\doinum{12.3456/s78910-011-012-3}
	%\artcitid{\#\#\#\#}
	%\volnum{123}
	%\year{2016}
%	\pgrange{1--8}
	\setcounter{page}{1}
%	\lp{8}
	% \tableofcontents
	% \listoffigures
	% \listoftables
	\section{Introduction}
	It has been more than a decade since the last super heavy elements (SHE) was discovered and named in honor of Russian scientist Yuri Tsolakovich Oganessian, "Oganesson" (Z=118) \cite{Z-118_ch5, Z_118_ch5}. A considerable amount of experimental and theoretical research is going on in the discovery of new SHE with Z $\geq$ 119. A newly built SHE factory \cite{DGFS_2022_ch5} at Joint Institute for Nuclear Research, Dubna, has already started carrying out experiments and publishing the results \cite{SHE_factory1_ch5, SHE_factory_ch5}. The recent experimental efforts to synthesize the SHE Z=119, 120 \cite{Z_119_ch5} were led by GSI researchers with gas-filled mass separator TASCA facility \cite{TASCA_ch5} where they had used $^{50}$Ti beam on $^{249}$Bk and $^{249}$Cf targets for over four months of beam time which resulted in zero evaporation residue events up to 6fb and 200 fb cross-section sensitivity for these two reactions, respectively. To successfully synthesize superheavy elements, it is imperative to have a thorough comprehension of the dynamics involved in heavy-ion induced reactions within the heavy-mass region. To gain a better understanding of the fusion-fission dynamics in heavy-ion induced nuclear reactions, extensive research has been conducted in recent decades. The investigation of experimentally measured evaporation residues is particularly significant in this context as it can provide valuable insights into the pre-saddle region of the compound nucleus. Therefore, the study of heavy-ion induced nuclear reactions can furnish critical information about the survival probability of compound nuclei, leading to an improved understanding of the formation of SHE. To study the dynamics of heavy elements, $^{240}$Cf was populated through $^{32}$S + $^{208}$Pb entrance channel for ER coss-section measurements. Earlier, Tsang \textit{et al}. \cite{Tsang1983_ch} have studied $^{32}$S-induced reactions on $^{208}$Pb to measure the fission fragment and angular distributions at 180, 210, 250, and 266 MeV laboratory energies. They also reported the distribution of folding angles between coincident fission fragments and concluded that the reaction proceeds \textit{via} the capture of the entire projectile by the target nucleus. Later, Back \textit{et al}. \cite{Back1985_ch5} reported angular distributions of fission fragments in $^{32}$S + $^{208}$Pb reaction. They found that the calculated cross sections are substantially larger than the experimental fission cross sections. This missing cross-section component may be associated with deep inelastic scattering, which was not included in the measurements of the symmetric mass component. Khuyagbaatar \textit{et al}. \cite{Khuyagbaatar2010_ch5, Khuyagbaatar2015_ch5} studied quasifission and fusion-fission in Cf isotopes. $^{242}$Cf was populated by using different isotopes of sulphur ($^{34}$S, $^{36}$S) and lead ($^{208}$Pb, $^{206}$Pb). The $^{34}$S + $^{208}$Pb neutron evaporation residue cross sections are found to be smaller by orders of magnitude than expected values for the $^{36}$S + $^{206}$Pb yields, assuming both reactions proceed by fusion followed by compound nucleus fission. The difference was explained by a smaller quasifission probability and a consequently larger probability of forming a compact compound nucleus in the $^{36}$S + $^{206}$Pb reaction. Nasirov \textit{et al}. \cite{Nasirov2019_ch5} showed the entrance channel effects in a heavy (S-Pb) system through measurements of the evaporation residues in $^{34}$S + $^{208}$Pb and $^{36}$S + $^{206}$Pb reactions. The difference was observed between the cross-section of evaporation residue formed in the '2n' and '3n' channels in the $^{34}$S + $^{208}$Pb and $^{36}$S + $^{206}$Pb reactions. Entrance channel characteristics, \textit{i.e.}, neutron and proton ratio of these reactions are the main reasons behind this observed difference. Hofman \textit{et al}. \cite{Hofman1994_ch5} obtained viscosity parameter of saddle-to-scission motion from measured giant dipole resonance $\gamma$ yield in the compound nucleus $^{240}$Cf. High energy $\gamma$ rays (E{$_\gamma$} {$\sim$} 20 MeV) were measured in coincidence with fission fragments from the reaction $^{32}$S + $^{208}$Pb ${\rightarrow}$  $ ^{240}$Cf at 200 and 230 MeV bombarding energies. Nuclear viscosity parameter $\gamma$ $\sim5$ was reported. Morton \textit{et al}. \cite{Morton1997_ch5} measured GDR $\gamma$ ray multiplicities for the $^{32}$S + $^{208}$Pb reaction in the energy range 180 - 245 MeV. These measurements demonstrated that the GDR decay mode is sensitive to the breakdown of the dynamical fission timescale. Previous studies \cite{Dioszegi2000_ch5,Shaw2000_ch5} calculated the nuclear viscosity parameter for the heavy $^{224} $Th and $^{240}$Cf compound nucleus systems. Similarly, $^{240}$Cf was populated using $^{32}$S beam and $^{208}$Pb target and $^{224} $Th was populated using $^{16}$O beam and $^{208}$Pb target. Dioszegi \textit{et al.} \cite{Dioszegi2000_ch5} and Shaw \textit{et al.} \cite{Shaw2000_ch5} have shown that the nuclear viscosity ($\gamma$) parameter obtained from the GDR data over a wide range of excitation energies does not reveal any temperature dependence. A complete review of the nuclear dissipation was done by Mazumdar \cite{MAZUMDAR2015_ch5}. Evaporation residues of $^{32}$S + $^{208}$Pb reaction system have not been measured so far, keeping this in view, it was planned to study the $^{240}$Cf nucleus in the present work using $^{32}$S + $^{208}$Pb reaction system. 
	\par \par The paper is organized into four sections. Section \ref{sec:level2} provides the experimental details followed by a detailed description of the data analysis in Section \ref{data_analysis} The sub-sections \ref{HYRA_eff1} and \ref{results} discuss the determination of HYRA efficiency, and experimental results of ER cross-sections, respectively, with summary in Section \ref{summary}

	\section{\label{sec:level2}Experimental Details}
\begin{figure}[h!]
		\centering
		\includegraphics[width=8.5cm]{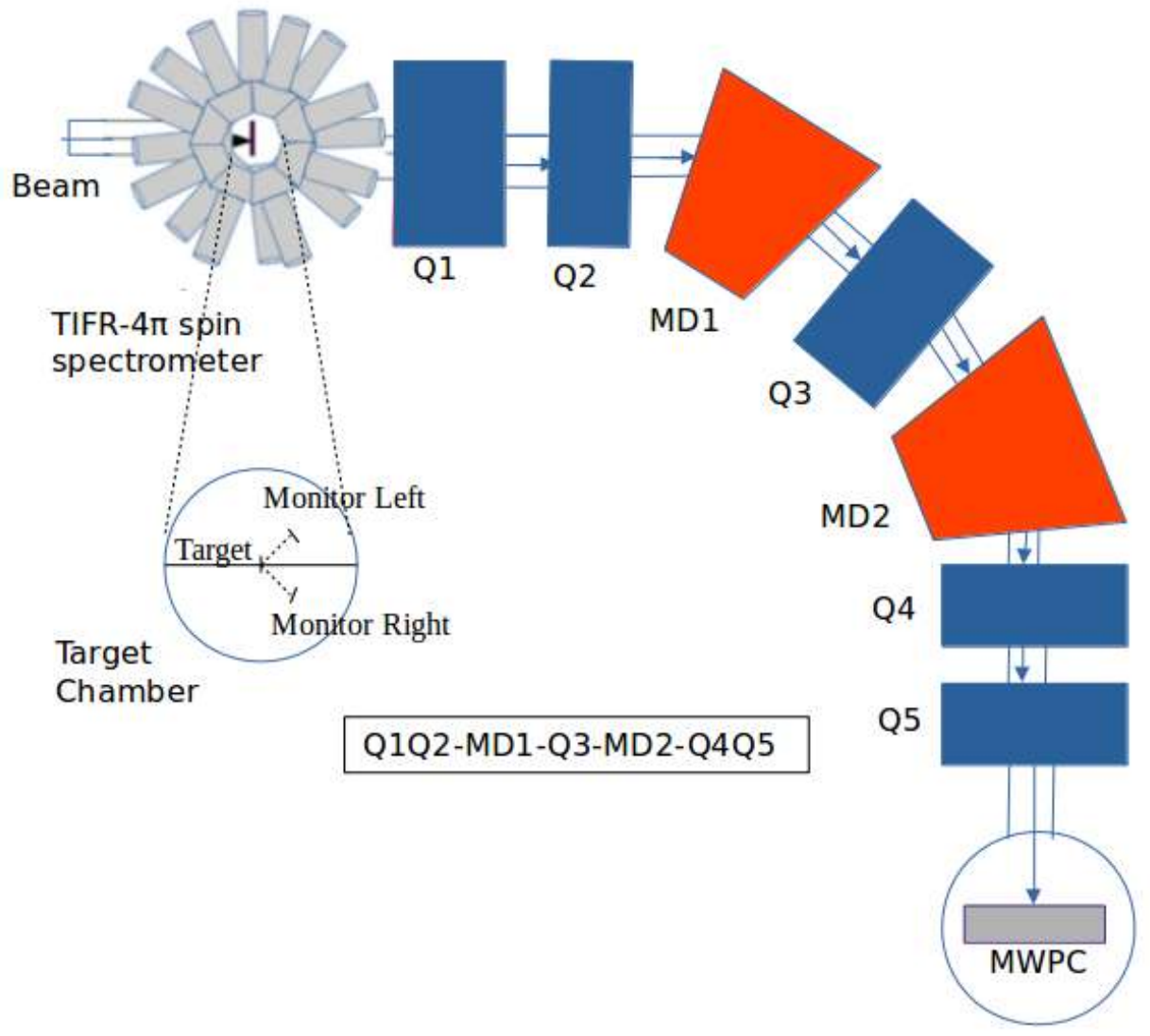}   
		% \caption{(Color online) A schematic diagram of TIFR 4$\pi$ Sum-Spin spectrometer and HYbrid Recoil Mass Analyzer (HYRA). Q1-Q5 are the magnetic quadrupole; MD1 and MD2 are magnetic dipoles. MWPC is the Multi-Wire Proportional Counter detector at the focal plane of HYRA.}
		\caption{(Color online) A schematic diagram of TIFR 4$\pi$ Sum-Spin spectrometer and HYbrid Recoil Mass Analyzer (HYRA). Q1-Q5 are the magnetic quadrupoles;
			MD1 and MD2 are magnetic dipoles. MWPC is the Multi-Wire Proportional Counter at the focal plane of HYRA.}
		\label{fig_1}
	\end{figure}

\begin{figure}[h!]
		\centering
		\includegraphics[width=8cm]{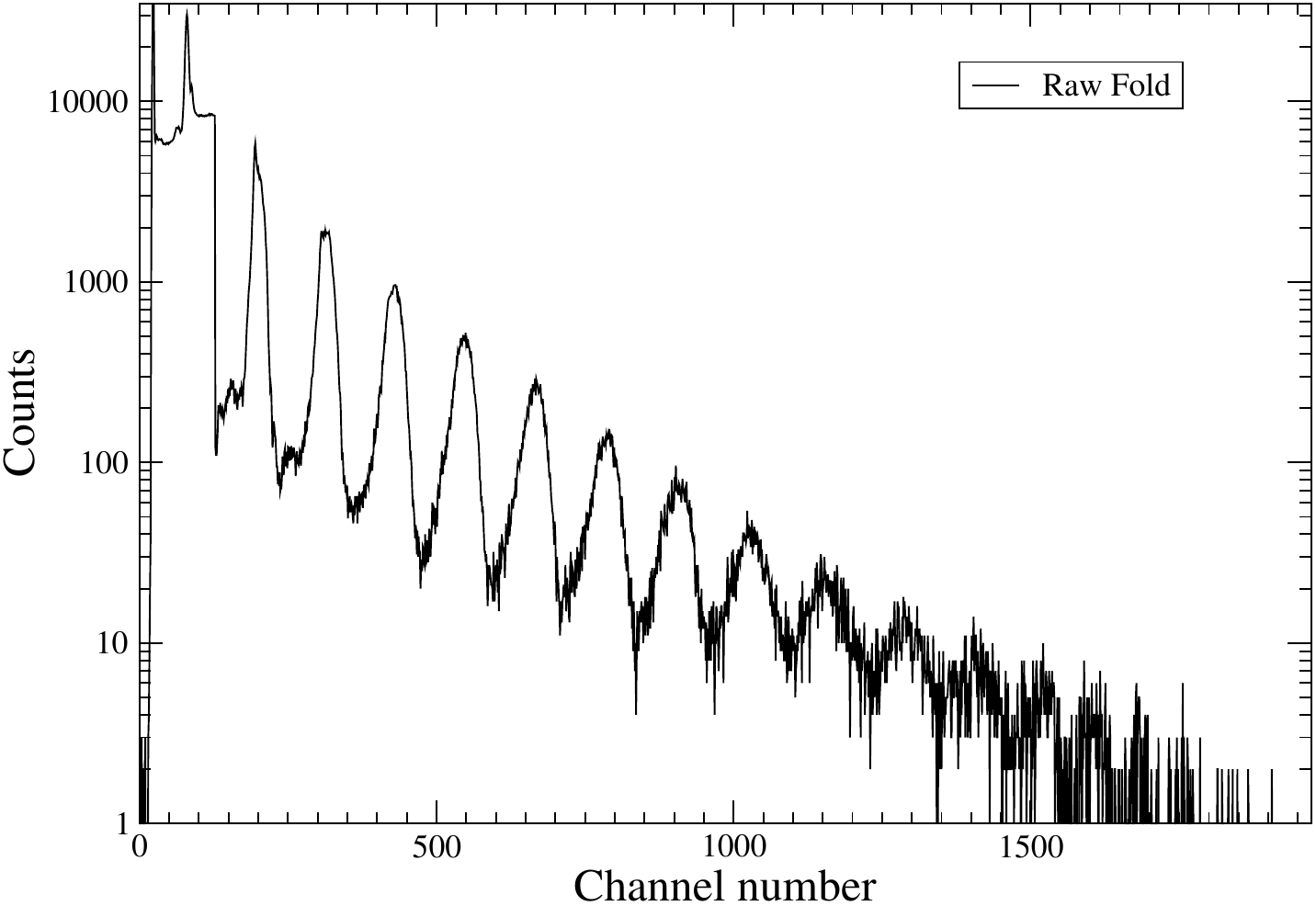}  
		\caption{Raw fold spectrum of $^{32}$S + $^{208}$Pb at lab energy 176.4 MeV.}
		\label{fig_2}
	\end{figure}

	%\cite{iuac_2022}
	The experiment was performed at the Inter-University Accelerator Center (IUAC), New Delhi \cite{iuac}. The $^{240}$Cf compound nucleus was populated by bombarding the $^{208}$Pb target with a pulsed beam of $^{32}$S from the Pelletron - LINAC accelerator facility at IUAC. An average beam current of $\sim$ 0.5 pnA was maintained throughout the experiment. The 97.9\% enriched \% isotopically enriched $^{208}$Pb target of thickness 227 $\mu$gm/cm$^{2}$ with carbon capping and backing of 25 $\mu$gm/cm$^{2}$ and 10 $\mu$gm/cm$^{2}$, respectively. The measurements were performed at lab energies of 176.2, 181.3, 186.4, and 191.5 MeV with pulse separation of 2 $\mu$s. The detection facility involved the IUAC HYbrid Recoil Mass Analyzer (HYRA) in gas-filled mode, coupled with the TIFR 4$\pi$ Sum-Spin spectrometer. The configuration of the electromagnetic elements of HYRA is Q1Q2-MD1-Q3-MD2-Q4Q5 where Q and MD stand for quadrupole and magnetic dipole, respectively. The schematic diagram of the setup is shown in Fig. \ref{fig_1}. For operating HYRA in gas-filled mode, He gas was used at a pressure of 0.21 Torr. A carbon foil with a thickness of 650 $\mu$g/cm$^2$ was used to segregate the gas-filled zone from the region under vacuum. For a fuller description and technical details of HYRA, we refer to \cite{madh_2010}. The optimization of the magnetic field values for different beam energies was done using a simulation program \cite{nath_mag}. The optimum field values were chosen by scanning around $\pm$ 10\% of the calculated values in steps of 2\% for each lab energy.  The Evaporation residues, filtered through the mass analyzer, were detected in a multi-wire proportional counter (MWPC) at the focal plane of HYRA. The MWPC had an active area of 150 mm × 50 mm and was operated using isobutane gas at a pressure of $\sim$ 2.0 Torr. A mylar foil was used to separate the MWPC from the gas-filled region of HYRA. Two silicon surface barrier detectors (referred to as monitor detectors in the text) were used to measure the Rutherford (elastically) scattered beam for normalizing the reaction cross sections. These detectors were placed inside the target chamber at a distance of 47 mm from the center of the target and at an angle of $\pm$ 23.4$^{\circ}$ with respect to the beam direction. HYRA is coupled with a TIFR-4$\pi$ spin-spectrometer, which is an array of 32 NaI(Tl) detectors in a soccer ball geometry covering 4$\pi$ solid angle to detect the low energy, discrete $\gamma$-rays. The NaI(Tl) detectors are conical in shape with 20 pentagonal and 12 hexagonal cross sections. Each of the detectors was energy calibrated with $\gamma$-ray sources of $^{137}$Cs (662 keV) and $^{60}$Co (1173 and 1332 keV). The cleanly separated ER events were obtained using the time-of-flight (TOF) approach. We used two time-to-amplitude converters (TACs) for this purpose. The first TAC measured the time difference between the start signal from the MWPC anode and the stop signal from the rf signal of the beam. The second TAC measured the time difference between the start signal from the MWPC-anode and the stop signal from logic OR of all the NaI(Tl) detectors in the 4$\pi$ spin spectrometer. The master strobe for data acquisition was the logic OR signal of the two monitor detectors and the anode signal of the MWPC. Fig. \ref{fig_2} shows the raw fold spectrum of $^{32}$S + $^{208}$Pb at lab energy 176.4 MeV. However, due to very low count rates, ER-gated spin distribution of the $^{240}$Cf cannot be measured.
	
	\section{\label{data_analysis}Data Analysis}
	This section presents a detailed account of the data analysis conducted for this experimental work. The CANDLE software was used for data collection and reduction \cite{et_candle}. In particular, Subsection \ref{HYRA_eff1} describes the procedure for calculating HYRA efficiency, which is critical for estimating ER cross-sections, and subsection \ref{results} presents the experimental results of ER cross-sections.
	
	\subsection{Determination of HYRA efficiency}\label{HYRA_eff1}
\begin{figure}[h!]
		\centering
		\includegraphics[width=8cm]{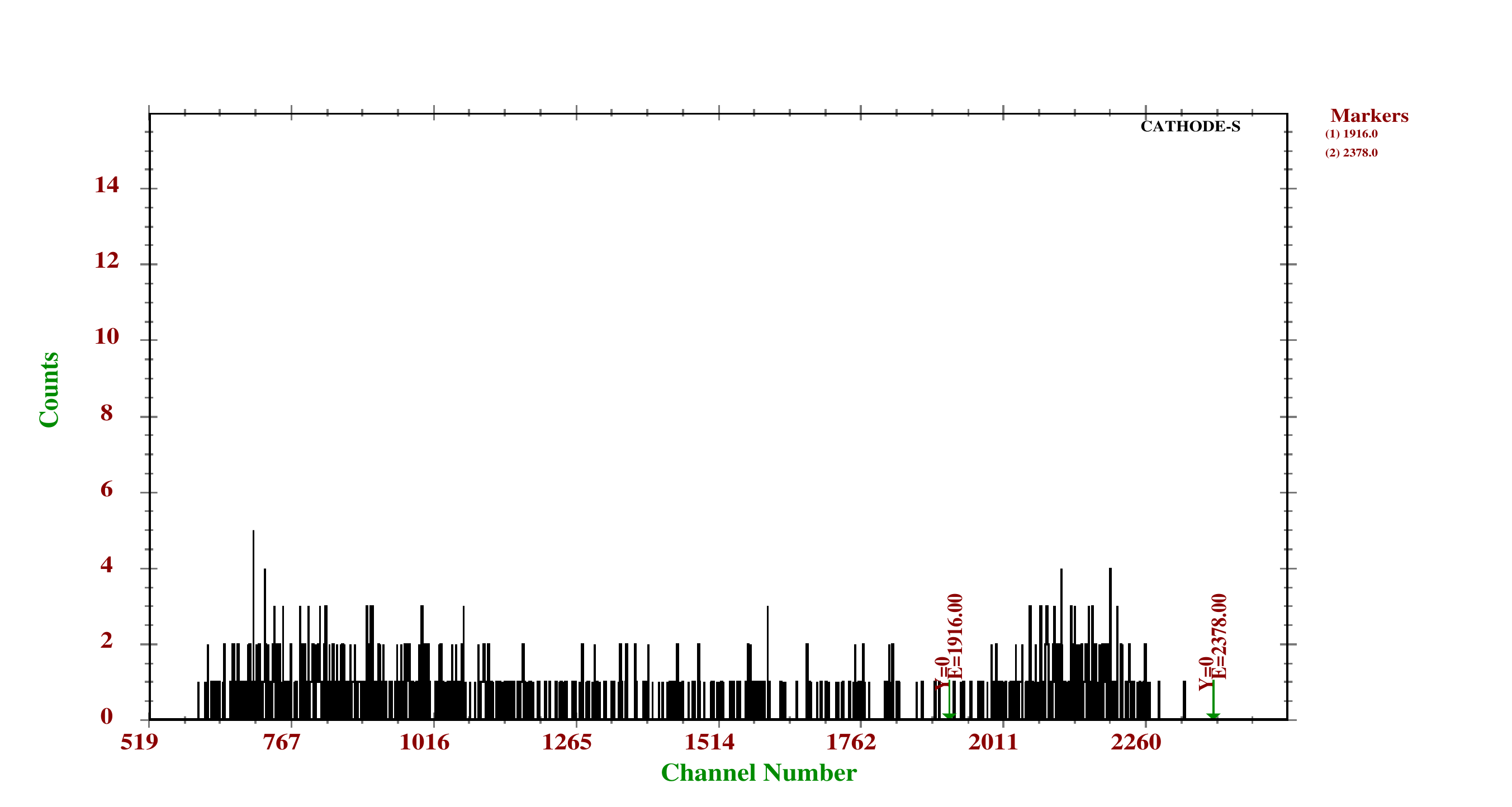} 
		\caption{Counts on the focal plane detector for $^{32}$S + $^{208}$Pb at 176.4 MeV lab energy.}
		\label{fig_4}
	\end{figure}

\begin{figure}[h!]
		\centering
		\includegraphics[width=8cm]{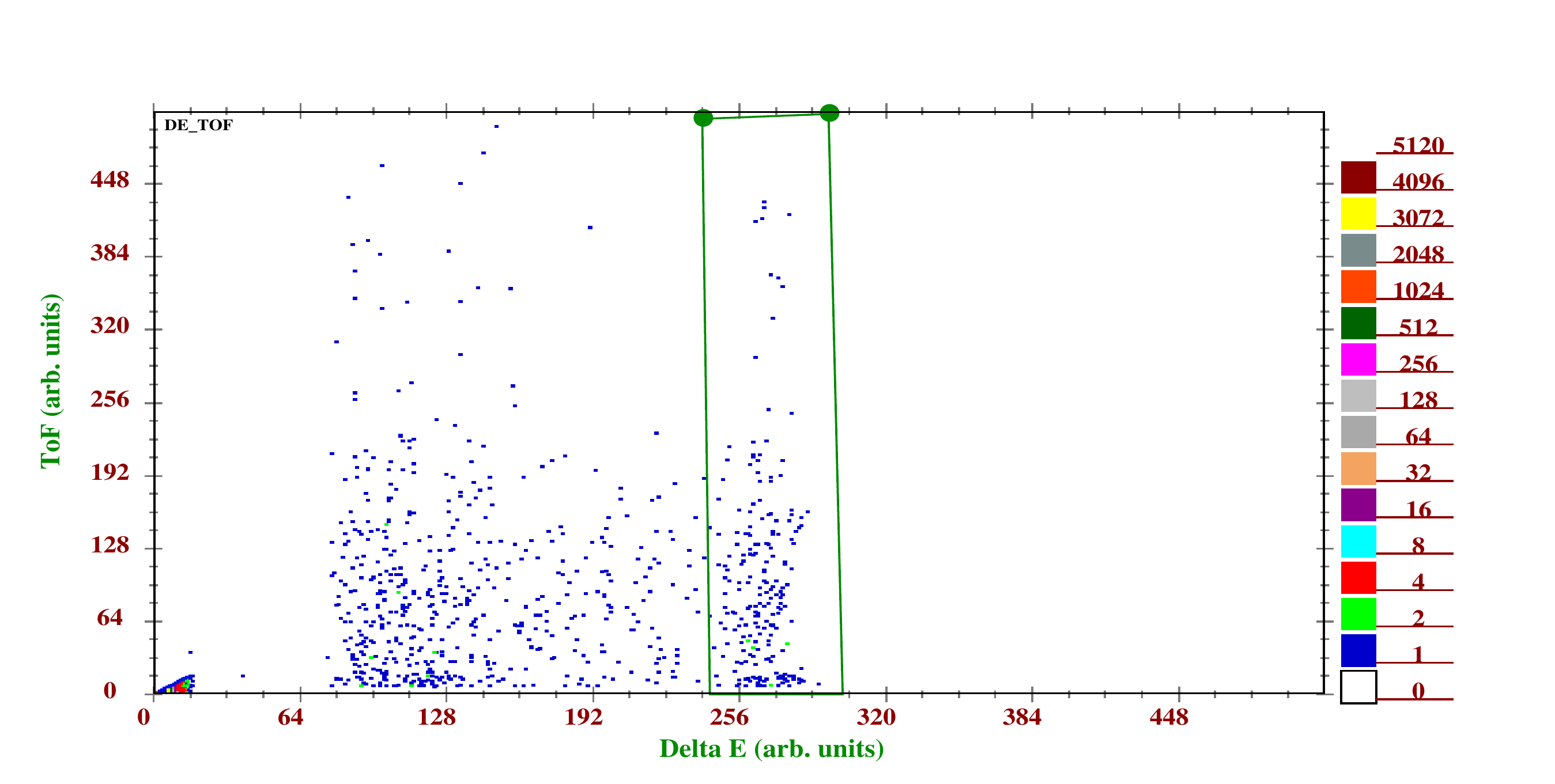}     
		\caption{Two-dimensional spectrum using  MWPC cathode signal and TOF spectrum of $^{32}$S + $^{208}$Pb at lab energy 176.4 MeV.}
		\label{fig_5}
	\end{figure}

\begin{figure}[h!]
		\centering
		\includegraphics[width=8cm]{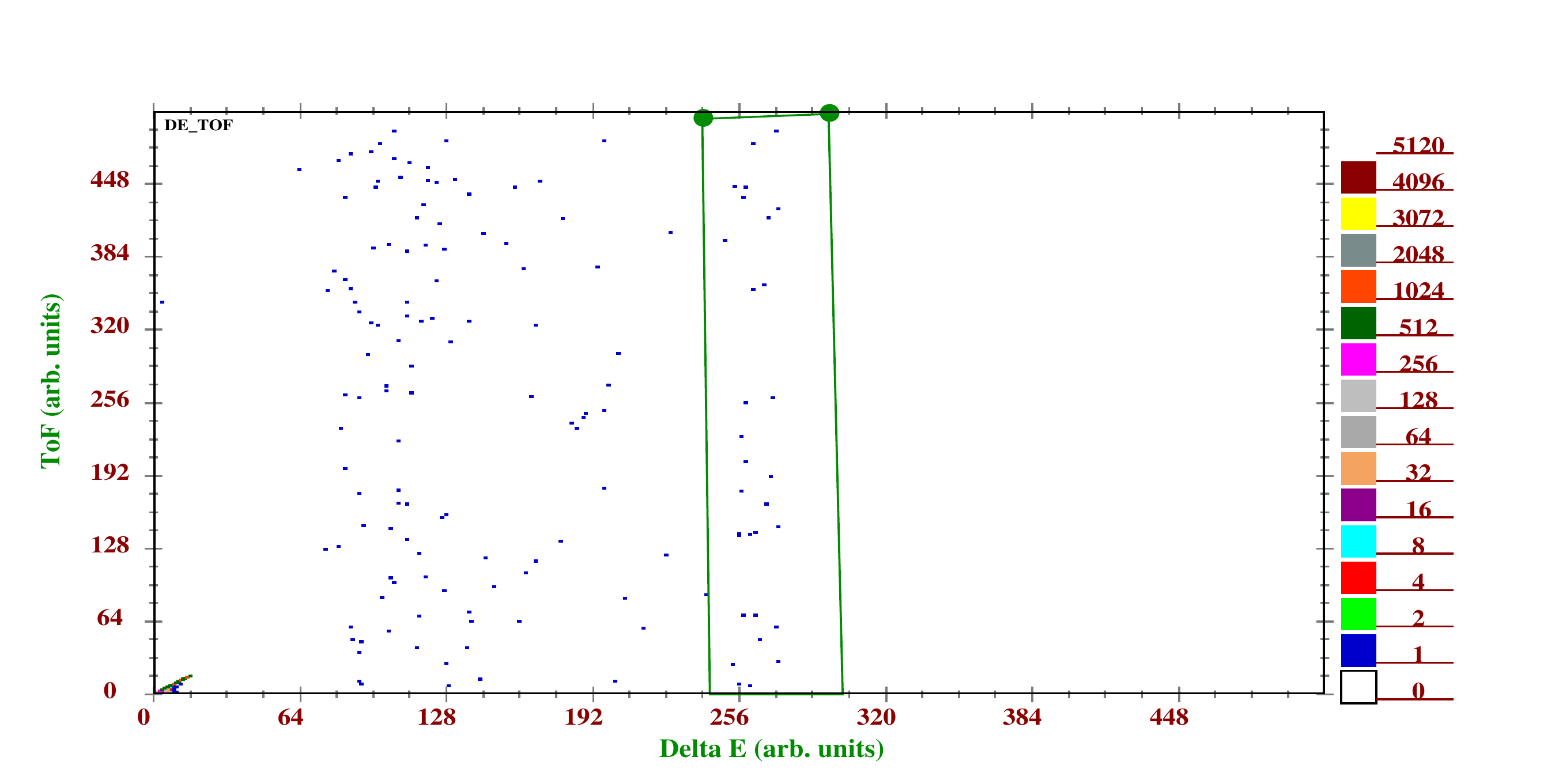}     
		\caption{Two-dimensional background spectrum using  MWPC cathode signal and TOF spectrum.}
		\label{fig_6}
	\end{figure}

	The transmission efficiency of HYRA is one of the important factors for the determination of the ER cross-sections. It can be defined as the ratio of the total ERs detected at the focal plane  to the total ERs produced in the target and depends on the beam energy, entrance channel, target thickness, exit  channel, magnetic field values, the angular acceptance of HYRA, gas pressure settings of the HYRA, and the size of the focal plane detector (MWPC) \cite{snath_2010}. To find the transmission efficiency (${\epsilon_{H}}$) for the present measurement, we have used the formula:
	\begin{equation}
		\epsilon_{H} = \frac{Y_{ER}}{Y_{Mon}} 
		\left( \frac{d\sigma}{d\Omega}\right)_{Ruth}\Omega_{Mon}\frac{1}{\sigma_{ER}}
		\label{equation_1}
	\end{equation}
	where $Y_{ER}$ is evaporation residue yield at the focal plane, $Y_{Mon}$ is the yield of elastically scattered projectiles detected by silicon surface barrier detectors and given by the geometric mean ($\sqrt{(Y_L Y_R)}$) of two detectors placed at $\pm$ 23.4$^{\circ}$ from the center of the target, $\left( \frac{d\sigma}{d\Omega}\right)_{Ruth}$ is the differential Rutherford scattering cross-section, ${\rm {\Omega_{Mon}}}$ is the solid angle subtended by monitor detector at the center of the target, ${\sigma_{ER}}$ is ER cross-section. The differential Rutherford scattering is given by, 
	\begin{equation}
		\left( \frac{d\sigma}{d\Omega}\right)_{Ruth} = 1.296 \left(\frac{Z_p Z_t}{E_{lab}}      \right)^{2}\left[\frac{1}{sin^4{(\frac{\theta}{2})}}- 2\left(\frac{A_p}{A_t}\right)^2\right]
	\end{equation} where $Z_p$, $Z_t$, and $A_p$, $A_t$ are the atomic and mass numbers of the projectile and target, respectively. $E_{lab}$ and $\theta$ are the energy of the incident projectile and scattering angle of the projectile-like particles in the laboratory frame of reference, respectively. \begin{figure}[h!]
		\centering
		\includegraphics[width=8cm] {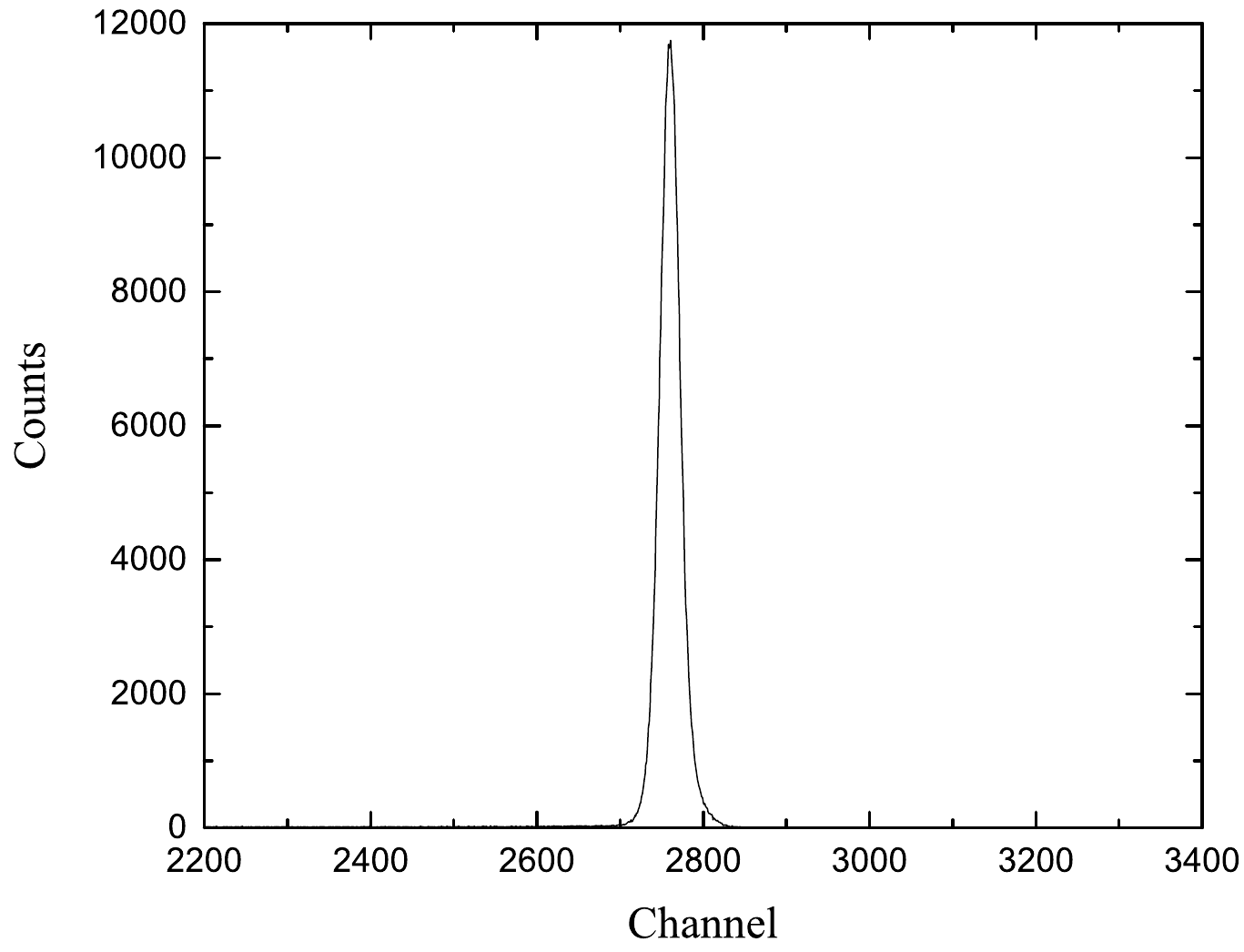}    
		\caption{Monitor detector spectrum kept on left at lab energy 176.4 MeV.}
		\label{fig_3}
	\end{figure}
	Fig. \ref{fig_3} illustrates the observed spectrum on the monitor detector positioned on the left side within the target chamber, utilizing laboratory energy of 176.4 MeV. On the other hand, Fig. \ref{fig_4} displays the spectrum recorded in the cathode of MWPC (Multi-Wire Proportional Chamber) at the same laboratory energy (E$_{lab}$=176.4 MeV). Only counts falling within the channel numbers 1916 to 2378 are taken into consideration. From our experience we know that counts in this region correspond to ERs from compound nuclei and a prominent peak would have appeared if the beam run was taken for a long time while a peak-like structure on the left side of the figure corresponds to the beam-like particles. Fig. \ref{fig_5} shows the 2D-spectrum showing ER of $^{32}$S + $^{208}$Pb at lab energy 176.4 MeV, and the background is shown in Fig. \ref{fig_6}. The counts in the gated region (shown in a green box) are considered for the cross-section calculation after subtraction from the background.\\        
	The efficiency of HYRA plays a major role in extracting ER-cross-section from experimental data. It is also the main source of error in the cross-section. In all the compound nuclei systems studied so far by the HYRA, the efficiency is calculated by the scaling method. This method is basically populating two different compound nuclei of similar mass with the same beam at the same energy, one for which ER cross-sections are already available in the literature, and the other is the main nucleus for which ER cross-sections are unknown. Using the known cross-section value from the literature, and experimental data we find the efficiency using Eq. \ref{equation_1}. Once this efficiency value is known, the main system efficiency is calculated by scaling the ER-angular distribution of the unknown system to the known system. ER-angular distribution is generated using the TERS code \cite{s.nath2_ch5, s.nath3_ch5, s.nath4_ch5}. For more description of efficiency calculation, we refer to \cite{ vsingh_2014_ch5, prasad_2001_ch5, r.sandal_2015_ch5}.
	\par
\begin{figure}[h!]
		\centering
		\includegraphics[width=8cm]{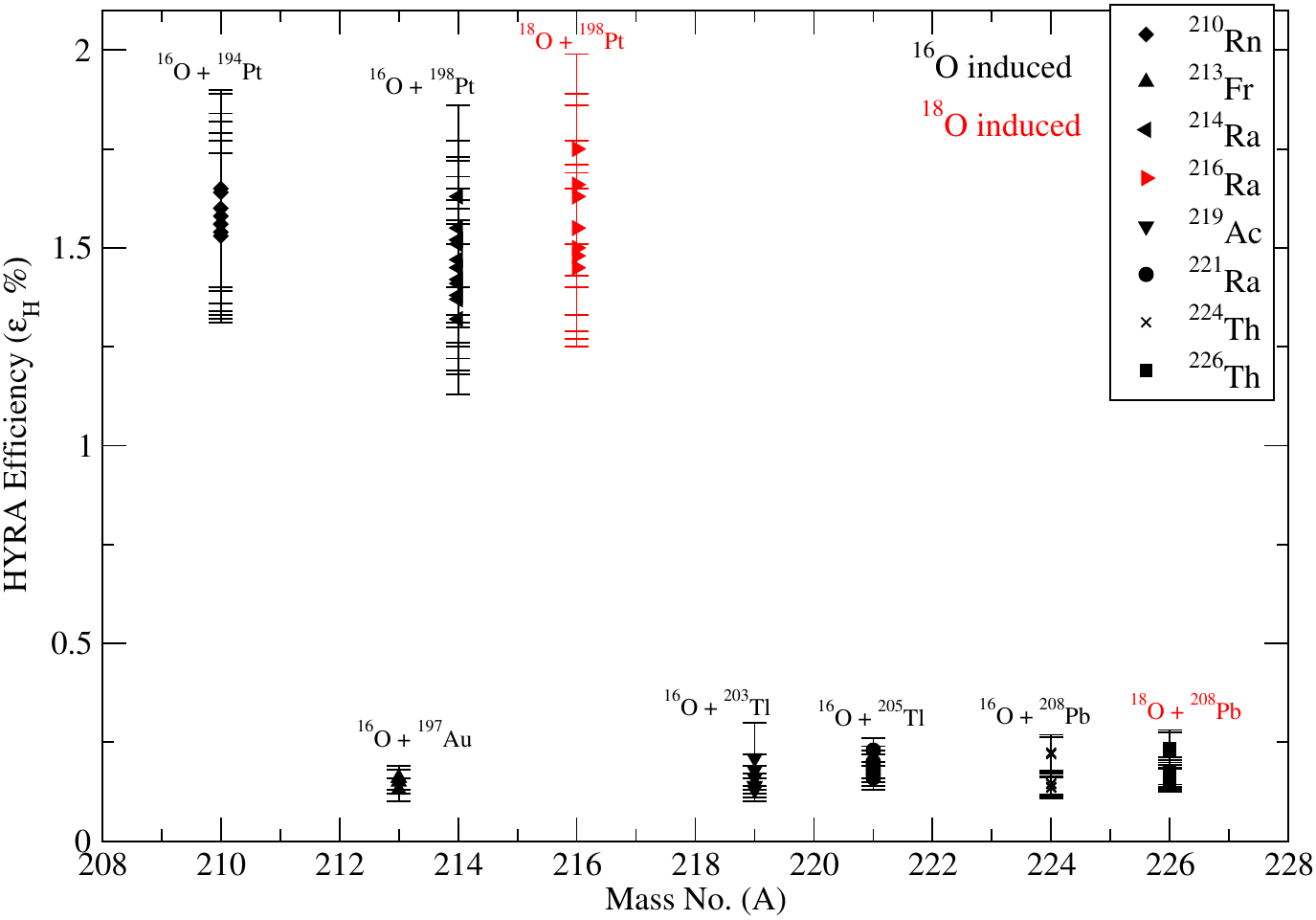}   
		\caption{Comparison of the HYRA efficiency ($\epsilon_H$) for $^{16,18}$Oinduced reaction as a function of mass number A at different lab energies.}
		\label{fig_7}
	\end{figure}

\begin{figure}[h!]
		\centering
		\includegraphics[width=8cm]{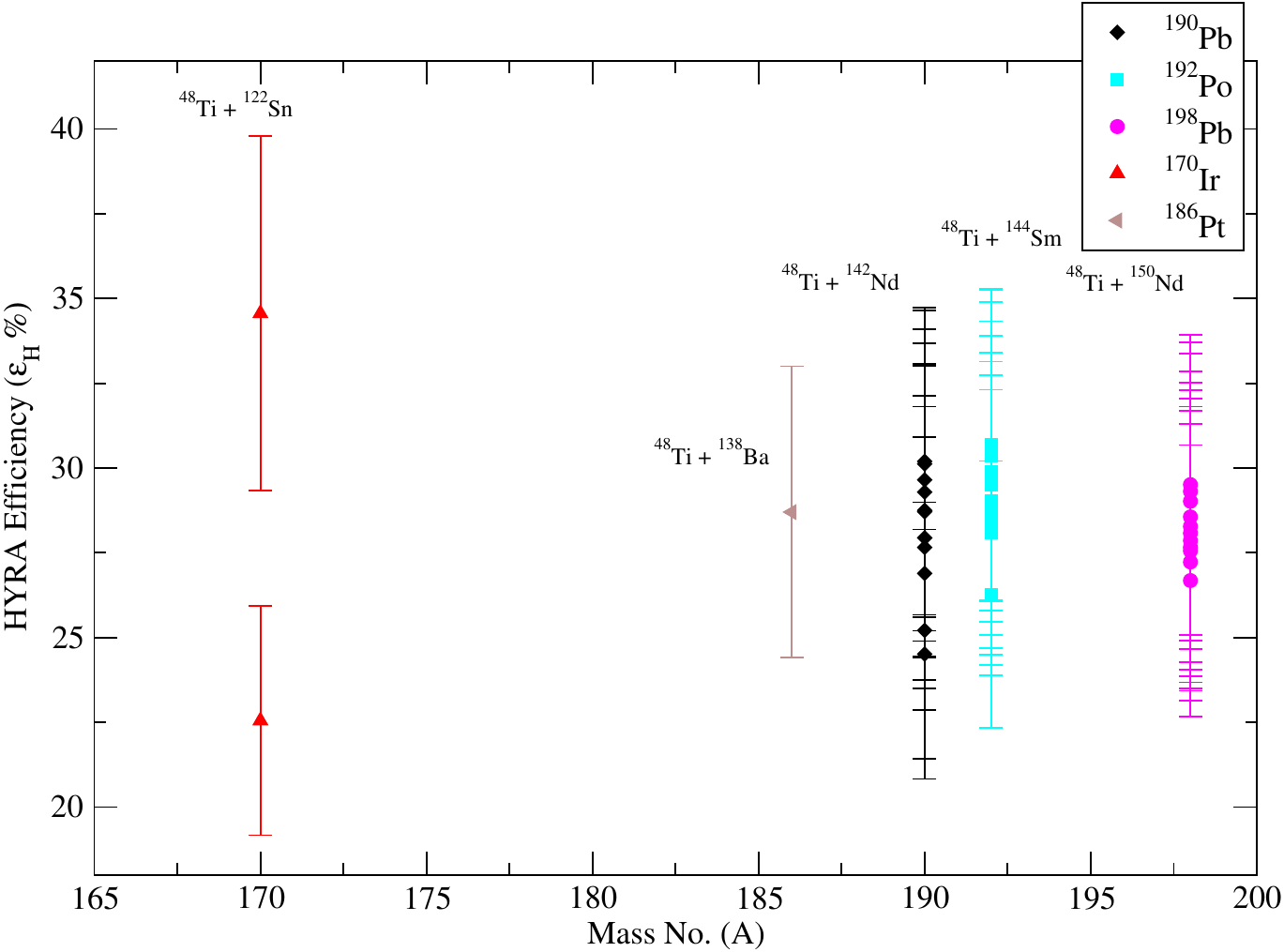}   
		\caption{Comparison of the HYRA efficiency ($\epsilon_H$) for $^{48}$Ti induced reaction as a function of mass number A at different lab energies.}
		\label{fig_8}
	\end{figure}

 But $^{32}$S + $^{208}$Pb $\rightarrow$ $^{240}$Cf, the compound nuclei case is different from the rest of the cases in the sense that it is in the heavy mass region and since it is so heavy the moment it forms it goes to fission and very less ER formation probability. Normally TERS code generates ER angular distribution for the decay channels predicted by the statistical model either PACE4 or DCASCADE \cite{cascade_2023}. But due to its very less ER formation probability, the statistical model cannot predict any decay channel for this system. So the traditional method of scaling can not be applied for HYRA efficiency calculation. But one can always give an estimate of the HYRA efficiency by studying the trend of efficiency in the heavy mass regions. A large range of masses is covered for the ER cross sections using HYRA. For example, the transmission efficiency of HYRA as a function of Mass number (A) for $^{16,18}$O induced reactions in the mass range 200-230 \cite{vsingh_2014_ch5,prasad_2001_ch5,r.sandal_2015_ch5,j.gehlot_2019_ch5,muthu_2020_ch5} is very wide as shown in Fig. \ref{fig_7}. At the same time, the transmission efficiency of HYRA as a function of Mass number (A) for $^{48}$Ti induced reactions in the mass range 170-200 \cite{priya_2017_ch5,rajesh_2019_ch5} is shown in Fig. \ref{fig_8}. There is a large spread of efficiencies for a given mass number.  This large spread is because these values are at different lab energies. HYRA efficiency of $^{48}$Ti-induced reactions lies in the range of 20 \% - 35 \% (Fig. \ref{fig_8}) and for $^{16,18}$O-induced reactions HYRA efficiency lies in the range of 0.03 \% - 1.75\% (Fig. \ref{fig_7}). \begin{figure}[h!]
		\centering
		\includegraphics[width=8cm]{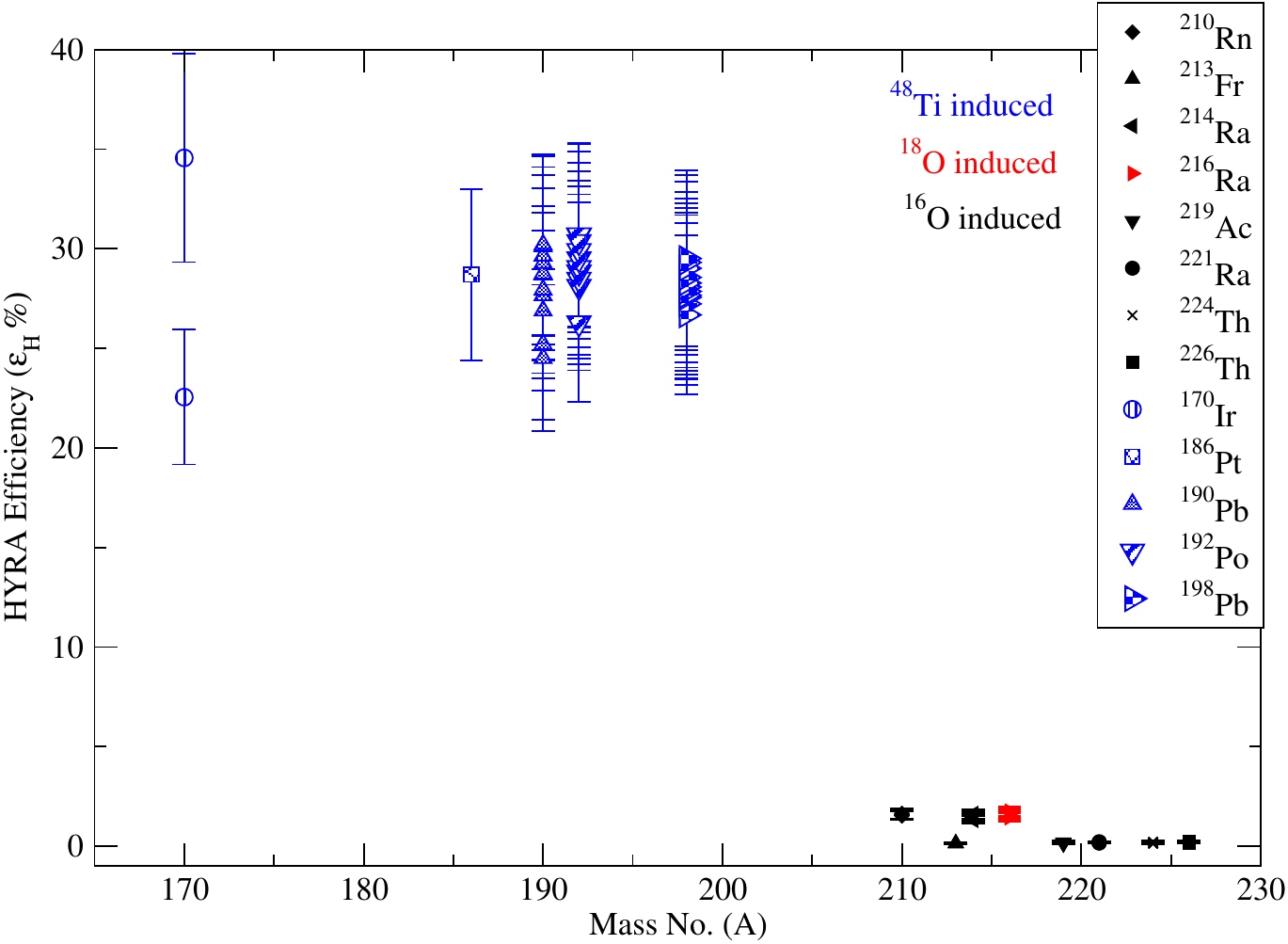}   
		\caption{The efficiency of HYRA ($\epsilon_H$) for $^{48}$Ti and $^{16,18}$O induced reaction as a function of mass number A at different lab energies.}
		\label{fig_9}
	\end{figure} For compound nucleus $^{213}$Fr, $^{219,221}$Ac and  $^{224,226}$Th HYRA efficiency lies within range of 0.13 \% - 0.224 \% \cite{j.gehlot_2019_ch5,muthu_2020_ch5} while for $^{210}$Rn and $^{214,216}$Ra HYRA efficiency lies within range of 1.32 \% - 1.75 \% \cite{prasad_2001_ch5,r.sandal_2015_ch5}. For $^{170}$Ir, $^{192}$Po, $^{190,198}$Pb HYRA efficiency lies within range of 22 \% - 35 \%. HYRA efficiency spread from mass number 170 to 226 is shown in Fig. \ref{fig_9}. From Fig. \ref{fig_9} it can be interpreted that for lower mass less than A $<$ 200, HYRA efficiency value is high whereas efficiency is very low in the heavy mass region (A $>$ 200).

	In a previous study \cite{Ranjan_ER}, the HYRA efficiency for the reaction $^{32}$S + $^{154}$Sm $\rightarrow$ $^{186}$Pt was calculated using the traditional scaling method. The determined efficiency fell within the range of 10\% to 11\%. The $^{186}$Pt system belongs to the A $<$ 200 mass region, which typically results in a higher HYRA transmission efficiency. For heavy ions such as $^{48}$Ti and $^{32}$S, the efficiency is higher when the mass region is A$<$200. But the efficiency for the $^{32}$S-induced reaction is lower than that for the $^{48}$Ti-induced reaction within the same mass region. Conversely, reactions induced by $^{16,18}$O in the A$>$200 mass region exhibit lower efficiency. 
	\par For $^{240}$Cf compound nuclei system which lies in A$>$200 mass region, it can be inferred that HYRA will have lower transmission efficiency but will be more than that of $^{16,18}$O induced reactions. So for the present ER cross-sections, HYRA transmission efficiency ($\epsilon_H$) range from 1 \% to 5 \% is used.
	\subsection{\label{results}Experimental results}
\begin{figure}[h!]
		\centering
		\includegraphics[width=8cm]{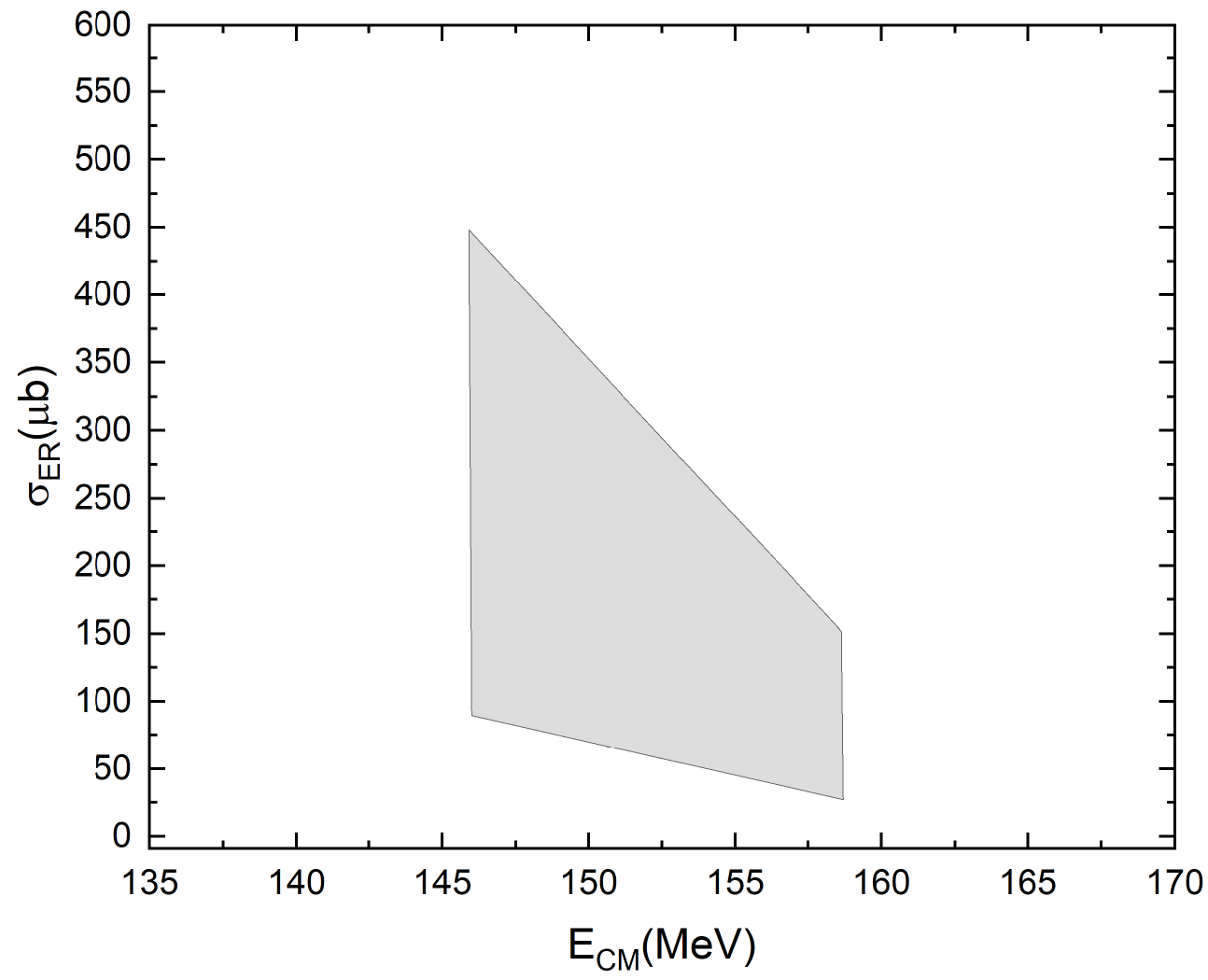} 
		\caption{Range of ER cross-section for $^{240}$Cf at different center of mass energies.}
		\label{fig_10}
	\end{figure}

	%%Use table* environment to get the table spanning both the columns
	Evaporation residue cross-sections were calculated using 
	\begin{equation}
		\sigma_{ER} = \frac{Y_{ER}}{Y_{Mon}} 
		\left( \frac{d\sigma}{d\Omega}\right)_{Ruth}\Omega_{Mon}\frac{1}{\epsilon_{H}}
		\label{equation_3}
	\end{equation}
	where symbols have the same meaning as that of equation \ref{equation_1}. The error in cross-section ($\sigma_{ER}$) comes from the error in $\epsilon_H$ and systematic error, and statistical error (obtained from the experimental parameters Y$_{ER}$ and Y$_{Mon}$). ER Cross-section at each energy was calculated using efficiency from 1\% to 5 \%. The experimentally obtained ER cross sections as a function of E$_{cm}$ for the $^{32}$S +$^{208}$Pb systems are shown in Fig. \ref{fig_10} and the shaded area in the figure shows the range of cross-section for this system. With the increase in the energy, a general trend of decreasing cross-section is observed. The results are tabulated in Table \ref{tableER240_ch5}. 
	\begin{table*}[htb]
%		\caption{Total ER cross-section of $^{32}$S +$^{208}$Pb at different energies.\\}
		\centering
		\label{tableER240_ch5}
		\vspace{0.2cm}
		%\begin{ruledtabular}
		\begin{tabular}{ l c c c c }
			%\cline{1-11
			%\renewcommand{\arraystretch}{1.2}
			\hline
			\vspace{0.1cm}
			E$_{lab}$(MeV) &  E$_{cm}$(MeV) & E$^{*}_{CN}$(MeV) & Hyra efficiency ($\epsilon_H$)$(\%)$ &$\sigma_{ER}$($\mu$b) \\[1ex] 
			%(MeV) & (MeV) & (MeV) & $(\%)$ &($\mu$b) \\[1ex]
			\hline 
			&  &  & 1  & 448     \\ [1ex]
			&  &  & 2  & 224     \\ [1ex]
			176.4 & 146.0 & 40.3 & 3  & 149     \\ [1ex]
			&  &  & 4  & 112    \\ [1ex]
			&  &  & 5  & 89      \\ [1ex]
			\hline
			&  &  &   1 & 334     \\ [1ex]
			&  &  & 2  & 167     \\ [1ex]
			181.3 & 150.1 & 44.4 & 3  & 111    \\ [1ex]
			&  &  & 4  & 84    \\ [1ex]
			&  &  & 5  & 66    \\ [1ex]
			\hline
			&  &  & 1  & 155       \\ [1ex]
			&  &  & 2  & 77    \\ [1ex]
			186.4 & 154.4 & 48.7 &   3  &  52   \\ [1ex]
			&  &  & 4  & 39   \\ [1ex]
			&  &  & 5  & 30    \\ [1ex]
			\hline
			&  &  & 1  & 151       \\ [1ex]
			&  &  & 2  & 75    \\ [1ex]
			191.5 & 158.6 & 52.9 &3  & 50    \\ [1ex]
			&  &  & 4  & 37    \\ [1ex]
			&  &  & 5  &30     \\ [1ex]
			\hline
		\end{tabular}
		%\end{ruledtabular}
	\end{table*}
	
	\section{\label{summary}Summary}Measurements of ER-cross-sections for $^{240}$Cf were conducted at four different energies using the $^{32}$S+$^{208}$Pb entrance channel. A general background run was taken over a long duration and used for background subtraction. The main source of uncertainty in the ER-cross-section measurement is the HYRA efficiency, which was extrapolated based on the study of all systems investigated at the HYRA facility. These experiments set a benchmark for measuring ERs in the same heavy mass regions (A$>$200) and represent one of the heaviest compound nuclei populated in India. These results put the upper limit on the measured cross-sections. Theoretical calculations for this system are currently underway and will be reported in another communication.
	\section*{Acknowledgments}
	One of the authors (R. Sariyal) acknowledges the Department Of Science \& Technology (DST), Govt. Of India for the INSPIRE fellowship. We thank the Pelletron and LINAC groups of IUAC for their support during the entire run of the experiment and for delivering quality beams. We also acknowledge the support of the target laboratory of IUAC.


\begin{thebibliography}{50}
		
		\bibitem{Z-118_ch5}{Yu. Ts. Oganessian, V. K. Utyonkov, Yu. V. Lobanov, F. Sh. Abdullin, A. N. Polyakov, R. N. Sagaidak, I. V. Shirokovsky,
			Yu. S. Tsyganov, A. A. Voinov, G. G. Gulbekian, S. L. Bogomolov, B. N. Gikal, A. N. Mezentsev, S. Iliev, V. G. Subbotin,
			A. M. Sukhov, K. Subotic, V. I. Zagrebaev, G. K. Vostokin, and M. G. Itkis, Phys. Rev. C \textbf{74}, 044602 (2006).}
		
		\bibitem{Z_118_ch5}{Yu. Ts. Oganessian \textit{et al.,} Phys. Rev. Let. 
			\textbf{109}, 162501 (2012).}
		
		
		
		
		\bibitem{DGFS_2022_ch5}Yu. Ts. Oganessian \textit{et al.,} Nucl. Instrum. Meth.
		Phys. Res. A \textbf{1033}, 166640 (2022).
		
		
		
		\bibitem{SHE_factory1_ch5}{Yu. Ts. Oganessian \textit{et al.,} Phys. Rev. C \textbf{106}, L031301 (2022).}
		
		\bibitem{SHE_factory_ch5}{Yu. Ts. Oganessian \textit{et al.,} Phys. Rev. C
			\textbf{106}, 064306 (2022).}
		
		
		
		
		\bibitem{Z_119_ch5} {J. Khuyagbaatar \textit{et al.,} Phys. Rev. C  \textbf{102}, 064602 (2020).}
		
		
		\bibitem{TASCA_ch5}{A. Semchenkov, W. Br\"{u}chle, E. J\"{a}ger, E. Schimpf, M. Sch\"{a}del, C. M\"{u}hle, F. Klos, A. T\"{u}rler, A. Yakushev, A. Belov, T. Belyakova, M. Kaparkova, V. Kukhtin, E. Lamzin and S. Sytchevsky, Nucl. Instrum. Meth. B \textbf{266}, 4153 (2008). }
		
		\bibitem{Tsang1983_ch} M. B. Tsang, D. Ardouin, C. K. Gelbke, W. G. Lynch, Z.R. Xu, B. B. Back, R. Betts, S. Saini, P. A. Baisden, and M. A. McMahan, Phys. Rev. C \textbf{28}, 747 (1983).
		\bibitem{Back1985_ch5}B. B. Back, R. R. Betts, J. E. Gindler, B. D. Wilkins, S. Saini, M. B. Tsang, C. K. Gelbke, W. G. Lynch, M. A. McMahan, and P. A. Baisden, Phys. Rev. C \textbf{32}, 195 (1985).
		
		
		\bibitem{Khuyagbaatar2010_ch5} J. Khuyagbaatar, F. P. Heßberger, S. Hofmann, D. Ackermann, V. S. Comas, S. Heinz, J. A. Heredia, B. Kindler, I. Kojouharov, B. Lommel, R. Mann, K. Nishio and A. Yakushev, The European Ph ysical Journal A \textbf{46}, 59 (2010).
		
		\bibitem{Khuyagbaatar2015_ch5}J. Khuyagbaatar, D. J. Hinde, I. P. Carter,  M. Dasgupta, Ch. E. Dullmann, M. Evers, D. H. Luong, R. du Rietz, A. Wakhle, E. Williams, and A. Yakushev, Phys. Rev. C \textbf{91}, 054608 (2015).
		
		
		\bibitem{Nasirov2019_ch5}A. K. Nasirov, B. M. Kayumov, G. Mandaglio, G. Giardina, K. Kim, and Y. Kim, The European Physical Journal A \textbf{55}, 29 (2019).
		\bibitem {Hofman1994_ch5} D. J. Hofman,  B. B. Back,  I. Dioszegi,  C. P. Montoya,  S. Schadmand,  R. Varma,  and P. Paul, Phys. Rev. Lett. \textbf{72}, 470  (1994).
		
		\bibitem{Morton1997_ch5} C. R. Morton, A. Buda, P. Paul, N. P. Shaw, J. R . Beene, N. Gan, M. L. Halbert, D. W. Stracener,  R. L. Varner, M. Thoennessen, P. Thirolf, and I. Dioszegi, Journal of Physics G \textbf{23}, 1383 (1997).
		
		
		\bibitem{Dioszegi2000_ch5}I. Dioszegi, N. P. Shaw, I. Mazumdar, A. Hatzikoutelis, and P. Paul, Phys. Rev. C \textbf{61}, 024613 (2000).
		
		\bibitem{Shaw2000_ch5}N. P. Shaw, I. Dioszegi, I. Mazumdar, A. Buda, C. R. Morton, J. Velkovska, J. R. Beene, D. W. Stracener, R. L. Varner, M. Thoennessen, and P. Paul, Phys. Rev. C \textbf{61}, 044612 (2000).
		
		\bibitem{MAZUMDAR2015_ch5}I. Mazumdar, Pramana \textbf{85}, 357 (2015).
		
		
		\bibitem{iuac}G. K. Mehta and A. P. Patro, Nucl. Instr. and Meth. A 268 (1988) 334.
		
		
		\bibitem{madh_2010}N. Madhavan, S. Nath, T. Varughese, J. Gehlot, A. Jhingan,
		P. Sugathan, A. K. Sinha, R. Singh, K. M. Varier, M. C.
		Radhakrishna, E. Prasad, S. Kalkal, G. Mohanto, J. J. Das,
		Rakesh Kumar, R. P. Singh, S. Muralithar, R. K. Bhowmik,
		A. Roy, R. Kumar, S. K. Suman, A. Mandal, T. S. Datta,
		J. Chacko, A. Choudhury, U. G. Naik, A. J. Malyadri, M.
		Archunan, J. Zacharias, S. Rao, M. Kumar, P. Barua, E. T.
		Subramanian, K. Rani, B. P. Ajith Kumar, and K. S. Golda,
		Pramana - J. Phys. 75 (2010) 317.
		
		
		\bibitem{nath_mag} S. Nath, A Monte Carlo code to model ion transport in dilute gas medium (unpublished).
		
		\bibitem{et_candle} E. T. Subramanium, B. P. Ajith Kumar, and R. K. Bhowmik,
		CANDLE: Collection and Analysis of Nuclear Data using Linux
		Network http://www.iuac.res.in/NIAS.
		
		\bibitem{snath_2010}S. Nath, P. V. M. Rao, S. Pal, J. Gehlot, E. Prasad, G. Mohanto, S. Kalkal, J. Sadhukhan, P. D. Shidling, K. S. Golda, A. Jhingan, N. Madhavan, S. Muralithar, and A. K. Sinha, Phys. Rev. C 81 (2010) 064601.
		
		\bibitem{s.nath2_ch5}  S. Nath, Nucl. Instr. Methods A \textbf{576}, 403 (2007).
		\bibitem{s.nath3_ch5}  S. Nath, Comput. Phys. Commun. \textbf{179}, 492 (2008). 
		\bibitem{s.nath4_ch5} S. Nath, Comput. Phys. Commun. \textbf{180}, 2392 (2009).
		
		
		\bibitem{vsingh_2014_ch5}  V. Singh,  B.  R. Behera,  M. Kaur,  A. Kumar, K. P. Singh,  N. Madhavan, S. Nath, J. Gehlot, G. Mohanto, A. Jhingan, I. Mukul, T. Varughese, J. Sadhukhan, S. Pal, S. Goyal, A. Saxena, S. Santra, and S. Kailas, Phys. Rev. C \textbf{89}, 024609 (2014).
		
		
		
		\bibitem{prasad_2001_ch5}E. Prasad, K. M. Varier, N. Madhavan,  S. Nath, J. Gehlot, S. Kalkal, J. Sadhukhan, G. Mohanto, P. Sugathan, A. Jhingan, B. R. S. Babu, T. Varughese, K. S. Golda, B. P. A. Kumar, B. Satheesh, S. Pal, R. Singh, A. K. Sinha, and S. Kailas, Phys. Rev. C \textbf{84}, 064606 (2011).
		
		\bibitem{r.sandal_2015_ch5}R. Sandal, B. R. Behera, V. Singh, M. Kaur, A. Kumar, G. Kaur, P. Sharma, N. Madhavan, S. Nath, J. Gehlot, A. Jhingan, K. S. Golda, H. Singh, S. Mandal, S. Verma, E. Prasad, K. M. Varier, A. M. Vinodkumar, A. Saxena, J. Sadhukhan, and S. Pal, Phys. Rev. C \textbf{91}, 044621 (2015).
		
		\bibitem{cascade_2023}R. Sariyal and I.Mazumdar, EPJ Web of Conferences \textbf{284}, 03022 (2023).
		
		\bibitem{j.gehlot_2019_ch5}J. Gehlot, A. M. Vinodkumar, N. Madhavan, S. Nath, A. Jhingan, T. Varughese, T. Banerjee, A. Shamlath, P. V. Laveen, M. Shareef, P. Jisha, P. S. Devi, G. N. Jyothi, M. M. Hosamani, I. Mazumdar, V. I. Chepigin, M. L. Chelnokov, A. V. Yeremin, A. K. Sinha, and B. R. S. Babu, Phys. Rev. C \textbf{99}, 034615 (2019).
		
		
		\bibitem{muthu_2020_ch5}M. M. Hosamani, N. M. Badiger, N. Madhavan, I. Mazumdar, S. Nath, J. Gehlot, A. K. Sinha, S. M. Patel, P. B. Chavan, T. Varughese, V. Srivastava, M. M. Shaikh, P. S. Devi, P. V. Laveen, A. Shamlath, M. Shareef, S. K. Duggi, P. V. M. Rao, G. Naga Jyothi, A. Tejaswi, P. N. Patil, A. Vinayak, K. K. Rajesh, A. Yadav, A. Parihari, R. Biswas, M. Dhibar, D. P. Kaur, M. R. Raju, and J. Joseph, Phys. Rev. C \textbf{101}, 014616 (2020).
		
		
		\bibitem{priya_2017_ch5}  P. Sharma, B. R. Behera, R. Mahajan, M. Thakur, G. Kaur, K. Kapoor,  K. Rani,  N. Madhavan, S. Nath, J. Gehlot, R. Dubey,  I. Mazumdar, S. M. Patel, M. Dhibar, M. M. Hosamani, Khushboo, N. Kumar, A. Shamlath, G. Mohanto, and S. Pal, Phys. Rev. C \textbf{96}, 034613 (2017).
		
		
		\bibitem{rajesh_2019_ch5}K. K. Rajesh, M. M. Musthafa, N. Madhavan, S. Nath, J. Gehlot, J. Sadhukhan, P. M. Aslam, P. T. Muhammed shan, E. Prasad, M. M. Hosamani, T. Varughese, A. Yadav, V. R. Sharma, V. Srivastava, Md. M. Shaikh, M. Shareef, A. Shamlath, and P. V. Laveen, Phys. Rev. C \textbf{100}, 044611 (2019).
		
		
		\bibitem{Ranjan_ER}R. Sariyal \textit{et. al.}, Phys. Rev. C (under communication)
	\end{thebibliography}
\end{document}